# Ultrafast opacity in borosilicate glass induced by picosecond bursts of laser-driven ions

B. Dromey[1], L. Stella[2], D. Adams[1], R. Prasad[1], K. F. Kakolee[1], R. Stefanuik[3], G Nersisyan[1], G. Sarri[1], M. Yeung[1], H. Ahmed[1], D. Doria[1], T. Dzelzainis[1], D. Jung[1], S. Kar[1], D. Marlow[1], L. Romagnani[1+], A. A. Correa[4], P. Dunne[3], J. Kohanoff[2], A. Schleife[4], M. Borghesi[1], F. Currell[1], D. Riley[1], M. Zepf[1*] and C. L. S. Lewis[1]

1. Centre for Plasma Physics, Department of Physics and Astronomy, Queens University Belfast, BT7 1NN, UK
2. Atomistic Simulation Centre, Department of Physics and Astronomy, Queens University Belfast, BT7 1NN, UK
3. School of Physics, University College Dublin, Belfield, Dublin 4, Ireland
4. Lawrence Livermore National Laboratory, Livermore, California, 94551, USA

*email: m.zepf@qub.ac.uk

+currently at Laboratoire LULI, Ecole Polytechnique

**Direct investigation of ion-induced dynamics in matter on picosecond (ps, $10^{-12}$ s) timescales has been precluded to date by the relatively long nanosecond (ns, $10^{-9}$ s) scale ion pulses typically provided by radiofrequency accelerators[1]. By contrast, laser-driven ion accelerators provide bursts of ps duration[2], but have yet to be applied to the study of ultrafast ion-induced transients in matter. We report on the evolution of an electron-hole plasma excited in borosilicate glass by such bursts. This is observed as an onset of opacity to synchronised optical probe radiation and is characterised by the $3.0 \pm 0.8$ ps ion pump rise-time ($\tau_P$). The observed decay-time ($\tau_D$) of $35 \pm 3$ ps i.e. $\tau_P \ll \tau_D$, is in excellent agreement with modelling and reveals the rapidly evolving electron temperature ($>10^3$K) and carrier number density ($>10^{17}$cm$^{-3}$). This result demonstrates that ps laser accelerated ion bursts are directly applicable to investigating the ultrafast response of matter to ion interactions and, in particular, to ultrafast pulsed ion radiolysis of water[3-5], the radiolytic decompositions of which underpin biological cell damage and hadrontherapy for cancer treatment[6].**

Whilst the dynamics of the early stages of ionising radiation interactions with matter have been investigated for photons[7] and electrons[8,9], the ultrafast interaction of high-energy ions with matter is fundamentally different[1]. A significantly higher linear energy transfer (LET, or stopping power $dE/dx$) for ions results in highly localized energy deposition along their trajectory in matter. The unambiguous characterization of the dynamics of a system immediately following such an interaction is of central importance to multiple fields in science and medicine such as hadrontherapy[6], the safety of astronauts[6], electronics for use in space[10] and nuclear engineering[11]. Therefore, the ability to generate sufficiently short pulses of ions is critical to advancing the understanding of ultrafast characteristics of ion interactions by direct experimental investigation.

To date ion-matter interactions have been studied using $\mu s$[12] and ns[1,13] ion pulses and have almost exclusively focused on radiolytic yields during pulsed radiolysis of aqueous solutions. Such studies are typically performed using chemical scavenging techniques[14,15] which themselves have the potential generate significant uncertainty at the concentrations required for high temporal resolution. In recent years, however, laser-driven ion sources have developed rapidly, and ion pulses with initial bunch durations < ps, high particle numbers (>$10^{13}$ per bunch) and excellent beam quality have been reported[2]. Such parameters provide an excellent basis for addressing the problem of time-resolved ion-matter interaction dynamics on ultrafast time scales for both aqueous solutions and amorphous/crystalline solids.

Here the excitation of electrons across the optical gap of a borosilicate (BK7) glass sample by a ps laser-driven ion burst is investigated. The resulting electron-hole plasma gives rise to transient opacity (TO) to optical probe radiation[7,8]. This TO is ideally suited to our study since it is expected a) to exhibit a growth time that is essentially instantaneous compared to our ion pulse duration for low doses[8,16], b) to subsequently relax rapidly on a timescale of 10's of ps[16] and c) does not consist of multiple pathways for species formation and decay, such as are found in the radiolysis of aqueous solutions[3-5]. We show how the pump and decay profiles of the TO are in excellent agreement with a semi-empirical Drude-like model (i.e. the charge carriers are assumed to be free and no selection rules are applied due to the lack of a crystalline structure) for our estimated ion pulse characteristics. This allows physical quantities that are otherwise extremely difficult to measure directly, such as the electron temperature and carrier density of the electron-hole plasma, to be estimated.

For this experiment, we used the TARANIS multi-TW laser facility at Queen's University Belfast[17] to accelerate $H^+$ ions (protons) that form part of a surface layer on a thin (10µm) gold target foil to an endpoint energy of ~15 MeV via the target normal sheath acceleration (TNSA) mechanism[2]. After spectral filtering in the irradiated sample we obtain an ion pulse of ~3 ps full width half maximum in the region of interest in the sample. The short ion burst duration is made possible by the duration of each energy slice (1 MeV bandwidth) ultimately being set by the initial ps pulse duration of the ion bunch[2] (see Methods).

The $H^+$ - glass interaction region was transversely probed using a synchronised ultrafast optical probe (1053 nm, 400 fs, $< 10 \, \mu J \, cm^{-2}$) from the main laser system to produce spatially resolved snapshots of the TO (Figure 1 a, b). The temporal evolution of the opacity is recorded by varying the relative time delay $\Delta t$ between the probe and ion beam. As shown in Figure 1 b), these snapshots reveal the spatial profile of the ion beam and also demonstrate the recovery of the initial state of the subject material as $\Delta t \rightarrow \infty$, indicating that no permanent macroscopic damage persists after the interaction.

However, obtaining a consistent data set of the temporal evolution of the TO in this manner is limited by shot to shot variation in $H^+$ flux (± 12 %) and spectral cut off energy (± 5 %). To permit a more accurate measurement of the TO, the full temporal evolution was instead recorded on a single shot basis. This was achieved by optically streaking[18,19] (Figure 1c) the region outlined by the red dashed boxes in Figure 1b).

In this arrangement the original ultrafast probe is stretched in time (or chirped) to create a long pulse (> 200 ps) with an approximately linear frequency sweep[18]. This pulse in turn permits direct mapping of the temporal evolution of the TO onto the spectrum of the probe pulse as it passes through the interaction region i.e. each successive frequency encodes a different level of opacity in the target. The optical streak is then achieved by passing the pulse over a diffraction grating to separate the frequency components spatially and imaging the interaction region onto a CCD[19]. In essence this allows the ultrafast dynamics of transient ion induced opacity to be observed along 1-dimension.

Optically streaked images of the TO for an increasing ion flux at the highest observed energies are shown in Figure 2 a-c), and are characterised by a corresponding increase in opacity at greater depth into the target. The maximum energy present in each shot can be extracted from the slope of the detectable onset of opacity (solid black line, Figure 2 b and c). A maximum energy of $15 \pm 0.75$ MeV was observed, confirming a sharp, $1 \pm 0.5$ MeV bandwidth cut-off in the TNSA $H^+$ spectrum. This observation is in line with spectral measurements made using radiochromic film (RCF) stacks. The temporal evolution of the data in Figure 2b) is examined in detail in Figure 3.

The principal feature of this optical streak is the reducing level, and greatly differing recovery time, of the TO induced in the target with respect to depth. Ion induced opacity is defined as a > 0.05 reduction in signal from full transmission ($1 \pm 0.03$). Another important feature is that the risetime of the TO is faster for traces taken at greater depths into the sample, corresponding to the progressively decreasing local duration of the broadband $H^+$ pulse due to spectral filtering.

The red trace in the main panel of Figure 3 b) shows how transmission at 1mm depth evolves as a result of a 3 ps full width half max impulse of $H^+$ ions (black trace Figure 3 b) delivering an estimated total dose of $60 \pm 15$ Gy in the region studied. Transmission reduces for the entire duration of the ion pump pulse (~10 ps). After this the medium is observed to rapidly recover with a time constant of 35 ± 3 ps ($^1/_e$ value of maximum opacity at 1mm depth). It is worth noting that due to their higher LET, protons create significantly more electron-hole pairs than a similar energy, dose and pulse duration electron beam in glass[7]. See Methods for pulse duration and dose estimation at 1mm depth.

To investigate the microscopic mechanism underlying the observed TO, we developed a simple Drude-type two-temperature model [20-25] for the relaxation of the electrons promoted to the conduction band after the passage of the H[+] ions in the glass (See Supplementary Infotmation). We assume that a uniform thermalized electron-hole plasma is quickly (< 0.1 ps) [16,24] generated inside the irradiated volume of the glass, the optical opacity of which provides a microscopic explanation of the observed macroscopic TO. For the ion pulse calculated in Figure 3 b) the model shows that approximately 5 ps after the initial excitation into the conduction band the electron-hole plasma can be considered to be uniform in the irradiated region and the temperature can be described by the electron temperature[24] (insets Figure 4). As can be seen in Figure 4, this provides an excellent fit to the experimental results discussed in Figure 3b) at 1 mm, and at 900 $\mu$m. The model suggests experimental peak carrier densities of 3-6 × $10^{17}$ cm[-3] and peak electron temperatures of ~ 3000 K An independent indication of the quality of our simple Drude-type model is given by the agreement of the estimated maximum of the stopping power (~ 9.7 eV/Å) with the *ab initio* time-dependent density-functional theory (TD-DFT) using the local adiabatic density approximation [27,28] estimate of ~ 10 eV/Å, which is within the experimental uncertainty of ± 10%.

While the model provides a good fit to the ion induced opacity at depths > 850 $\mu$m, where the total dose is moderate, this is not true for shallower depths in the target (i.e. < 850 $\mu$m where the transmission does not return to the initial value). For the higher doses at those shallower depths (> 100 Gy) initial investigations show that considering a radiative bottleneck to electron relaxation in our model increases the accuracy of the fit, but a study into electron-hole plasma evolution coupled with localised lattice heating and kinetic radiative bottleneck formation is beyond the scope of our study here and will form the basis for future work. In our discussion we are exclusively concerned with interactions where the pulse is sufficiently short and the dose sufficiently low such that lattice heating can be neglected.

In conclusion, we have observed ion-induced dynamics on picosecond timescales during high LET interactions with matter. The intrinsic synchronicity of the laser-driven ion and optical probe beams ensures a high degree of reproducibility in the optical streaking technique. The rise time of the transient opacity is observed to be on the order of the pump pulse duration and rapid recovery to the initial state, with an time constant of 35 ± 3 ps, is consistent with our Drude-like model for electron – hole plasma dynamics. Overall this permits direct estimation of the picosecond time-dependent response of electronic temperature and carrier density in matter to ultrafast pulses of ions. This technique opens the way for ps pulsed ion radiolysis of aqueous solutions, and in particular that of water[3-5] which is of specific relevance to hadrontherapy[6] and neutron radiolysis for nuclear engineering[11], to be performed at even small scale laboratories.

**Methods**

**Ion pulse characteristics and intrinsic pulse shortening:**

Substantial stretching of the initial broadband ion bunch as it propagates with a large velocity spread from the target towards the sample results in an ion pulse > 70 ps (since the glass sample is 5 mm behind the target). The ultimate ion pulse duration, however, is a function of depth of propagation into the target. This is due to spectral filtering resulting from continuous energy loss and stopping of lower energy ions in the target, so that by the end of the track (~1 mm into the target) the interaction is essentially a few ps, few MeV ion pulse, limited in essence by the original ultrashort pulse duration of the original interaction. The ion pulse temporal profile, energy bandwidth and flux at a given depth are estimated from semi-empirical calculations (SRIM[26]) of the ion stopping power in $SiO_2$ based on the experimentally observed $H^+$ spectral shape. These calculations are also in good agreement with an *ab initio* estimate of the ion stopping power in a similar material (cristobalite with experimental density imposed) obtained from TD-DFT[27,28] calculations, the full details of which are described in the Supplementary Information.

The ion pulse in Figure 3b (black trace) is calculated as follows: The Bragg induced cut-off (~118 ps) in the pulse shape relates to ions with initial energy of approximately $13.1 \pm 0.7$ MeV incident on the sample. Energies below this have already been stopped in the sample and do not contribute to the TO at this depth. The high energy cut-off (~108 ps) is provided by experimental observation showing a $15.0 \pm 1.5$ MeV fast ion front. This analysis demonstrates that the ion pulse driving the TO at 1mm depth is on the order of $3.0 \pm 0.8$ ps at the half maximum point, and that the total pumping time (ultimate duration of the pulse) is on the order of $10.0 \pm 2.5$ ps for this data set. These define start and endpoints of the pulse. The pulse profile is calculated as a convolution of the on-axis spectral shape of the ions coupled with the continuous energy loss due to propagation in the glass (corrected for straggling) using SRIM calculations. The residual ion energy remaining in this pulse taking into account the continuous energy loss is shown by the colour scale on the left of Figure 3b).

**Acknowledgements**

This work was supported by the Engineering and Physical Sciences Research Council (EPSRC) through both a Career Acceleration Fellowship (EP/H003592/1) and a platform grant (EP/I029206/1). R.S. was supported by Science foundation Ireland (08/RFP/Phy1180). A.A.C. and A.S. performed work under the auspices of the U.S. Department of Energy by Lawrence Livermore National Laboratory under Contract DE-AC52-07NA27344.

**Author contributions**

The concept of this experiment was developed by M. Z., C. L. S. L. and B.D. The experiment was performed in the main by R. P., R. S., D. A. and B.D. Modelling was performed by L.S., A. A. C., A. S. and J.K. All authors contributed directly to the general experiment, data analysis and discussion, and manuscript preparation.

**Figure 1. Probing of ultrafast ion induced opacity in a dielectric material.** Figure 1a) shows a schematic of the experimental setup. An intense laser pulse ($> 10^{19}$ Wcm$^{-2}$) is incident at an angle of 45° on a 10 $\mu$m Au foil. An ion pulse produced via the TNSA mechanism[2] in the direction normal to the rear surface of the foil travels a distance of 5 mm to an optical quality BK7 glass slide (1 mm thick in horizontal dimension). The horizontal extent of the ion pulse is apertured to $500 \pm 50$ $\mu$m using a 1mm thick Al collimating slit (CS), centred on the glass sample (Top view inset). Low energy ions, electrons and X-rays were filtered using a 5$\mu$m Al foil in front of the CS (not shown). A synchronised probe pulse that can be either ~ 400 fs (compressed) or ~ 210 ps (chirped) is used to transversely image the pumped region (red line) onto an optical Andor CDD. Figure 1 b) shows a series of probe (400 fs) images taken at different time delays $\Delta t$ with respect to the arrival of the main laser pulse at the Al foil. The different penetration depths transverse to the beam propagation axis are due to the radially varying peak ion energy typical of such laser-driven accelerators. The temporal evolution of ion induced opacity along the red dashed region in Figure 1b) on a single shot basis is shown in Figure 1c). This region ($100 \pm 20$ $\mu$m) was imaged onto the entrance slit of a 1m imaging spectrometer with the pulse duration set to ~ 210 ps. The temporal evolution of TO was then encoded as variation of the signal in the probe spectrum (bandwidth ~ 4nm) with respect to depth in the target. Figure 1c) shows a typical streaked image[18,19], as detected using an optical Andor CCD.

**Figure 2. Optical streaking of ultrafast ion induced kinetics in dielectric material.** In this series of streaks the maximum proton energy is $15 \pm 3$ MeV with ion flux increasing at the highest energies for each successive shot. As can be seen the depth of penetration of the ion beam rapidly increases, as evidenced by increasing TO at deeper depths. Another key feature of these streaked images is the observation of a very sharp ion front extending into the target (black solid lines represent 15 MeV, no stopping included). This is very clear evidence of a sharp cut-off in the TNSA spectrum with a bandwidth of $1 \pm 0.5$ MeV. The dashed horizontal line (2 b and 2 c) shows the expected spatio-temporal evolution of a 13 MeV ion front taking into account continuous energy loss in the medium as calculated using ab-initio TDDFT calculations[27,28] and SRIM[26]. It is this property that leads to ultrashort ion pulse generation. While H$^+$ with incident energy of 15 MeV will penetrate further than 1 mm (vertical dotted line), H$^+$ with incident energy of $13 \pm 0.3$ MeV will terminate fully due to energy loss in the medium (curved dashed line). H$^+$ with incident energy < 12.7 MeV will not reach 1 mm depth. This intrinsically formed picosecond interaction permits a new regime of linear energy transfer to be explored at deeper depths in the target – ultrafast ion interactions with matter and the resultant short lived species, in this case an electron hole plasma.

**Figure 3. Spatially resolved temporal evolution of transient ion induced opacity with picosecond scale ion pulses (part empirical, part analytically derived).** Figure 3a) shows line outs of transmission at the depths labelled from the image in Figure 2b) (displayed on the floor of the plot). Here we define the onset of TO for a reduction in transmission > 5% due to fluctuations of ± 3% on the spatial intensity profile probe beam. In Figure 3b) the dark red trace shows the evolution of opacity in the glass at a depth of 1 mm. The horizontal thick translucent red line shows the uncertainty in the transmission of the probe through the glass in the absence of the ion pump beam. The black trace shows the calculated ion pump pulse profile (normalised) taking into account the spectral shape (2 MeV exponential) and cumulative energy loss at a depth of 1 mm in the glass target (normalised for clarity of comparison with transmission data). The colour gradient under the curve (colour scale on left of Figure 3b) of shows the residual energy of the remaining ions as calculated using SRIM[26] and shows a spread of ~ 0 MeV to ~ 5 MeV. The inset of Figure 3 b) shows the lineouts in Figure 3 a) in a 2-D representation for direct comparison. It is critical to note that for shallow depths (< 850 $\mu$m) the opacity is heavily saturated and hence does not return reliable information on the transmission profile. The approximate pump ($\tau_P$) and decay ($\tau_D$) times are shown for illustrative purposes only.

**Figure 4. Drude-like model for electron-hole plasma excitation and relaxation fit to the experimental data at a time after the ion generation, $T_0$.** Fits (black solid and dashed lines) to the data (squares and circles) are show for the evolution of transmission at 1mm and 900 $\mu$m depths of penetration into the target (+ 5 mm for distance to ion source). The insets show the evolution of electron temperature, $T_e$, (bottom) and density, $N_e$, (top) for each depth according to the adjusted two temperature model; then these quantities are used to calculate the opacity based on the dielectric function of a Drude model for $N_e$ carriers. It is important to note that the model returns accurate values for ion stopping at depths > 850 $\mu$m where the transmission returns to its original value. At depths < 850 $\mu$m, where the transmission does not return to its original value (Figures 3 a, 3 b inset), the model does not return a good fit to the data, suggesting that a kinetic bottleneck due to increased dose is preventing full electron-hole plasma relaxation on the timescales studied here at these depths. See Supplementary Information for full detail of the Drude-type two temperature model, fitting procedure, and TDDFT[27,28] calculations used to model the experimental data.